\documentclass[%
 reprint,
 superscriptaddress,
 amsmath,amssymb,
 aps,
 prl,
]{revtex4-2}

\usepackage{graphicx}
\usepackage{dcolumn}
\usepackage{bm}

\usepackage{accents}

\usepackage{hyperref}

\usepackage{xcolor}

\usepackage{ulem}

\begin{document}

\title{Floquet engineering a bosonic Josephson junction}

\author{Si-Cong Ji}
\affiliation{ Vienna Center for Quantum Science and Technology (VCQ), Atominstitut, TU Wien, Vienna, Austria}
 
\author{Thomas Schweigler}%
\affiliation{ Vienna Center for Quantum Science and Technology (VCQ), Atominstitut, TU Wien, Vienna, Austria}
\affiliation{JILA, University of Colorado, Boulder, Colorado, USA}

\author{Mohammadamin Tajik}%
\author{Federica Cataldini}%
\affiliation{ Vienna Center for Quantum Science and Technology (VCQ), Atominstitut, TU Wien, Vienna, Austria}

\author{Jo\~{a}o Sabino}%
\affiliation{ Vienna Center for Quantum Science and Technology (VCQ), Atominstitut, TU Wien, Vienna, Austria}
\affiliation{Instituto Superior T\'{e}cnico, Universidade de Lisboa, Lisbon, Portugal}
\affiliation{Instituto de Telecomunica\c{c}\~{o}es, Physics of Information and Quantum Technologies Group, Lisbon, Portugal}

\author{Frederik S. M{\o}ller}%
\author{Sebastian Erne}%
\author{J\"{o}rg Schmiedmayer} \email{schmiedmayer@atomchip.org}
\affiliation{ Vienna Center for Quantum Science and Technology (VCQ), Atominstitut, TU Wien, Vienna, Austria}

\date{\today}

\begin{abstract}
We study Floquet engineering of the tunnel coupling between a pair of one-dimensional bosonic quasi-condensates in a tilted double-well potential. By modulating the energy difference between the two wells, we re-establish tunnel coupling and precisely control its amplitude and phase. This allows us to initiate coherence between two initially uncorrelated Bose gases and prepare different initial states in the emerging sine-Gordon Hamiltonian. We fully characterize the Floquet system and study the dependence of both equilibrium properties and relaxation on the modulation.
\end{abstract}

\maketitle

\textit{Introduction}.---Periodic driving, i.e.~Floquet engineering, offers a wide range of pathways to design effective Hamiltonians which are otherwise difficult to achieve or even unrealizable in static laboratory systems \cite{Wei21, Eck17, Gol15}. For ultracold atoms, Floquet engineering has been used in optical lattices to coherently control tunneling between sites \cite{Lig07,Kie08, Eck09,Aid13, Miy13}, applied to generate large artificial gauge fields \cite{Dal11, Str11,Str12} and to simulate the quantum Hall effect \cite{Stu15, Man15, Liu10, Wu16}. Proposals for continuous systems range from implementing the Pokrovsky-Talapov model \cite{Pok79, Laz09, Kas20}, used to describe the Commensurate-Incommensurate phase transition \cite{Bak82}, to analogue simulators for 
pre-heating \cite{Amin15} and false-vacuum decay \cite{Col77}. Experimental studies on continuous interacting many-body Floquet systems and their equilibration process are, however, limited. In this letter we present an experimental study of Floquet engineering tunneling and phase locking in a pair of continuous one-dimensional superfluids in a tilted double well (DW).

\begin{figure}
\includegraphics[scale=0.21]{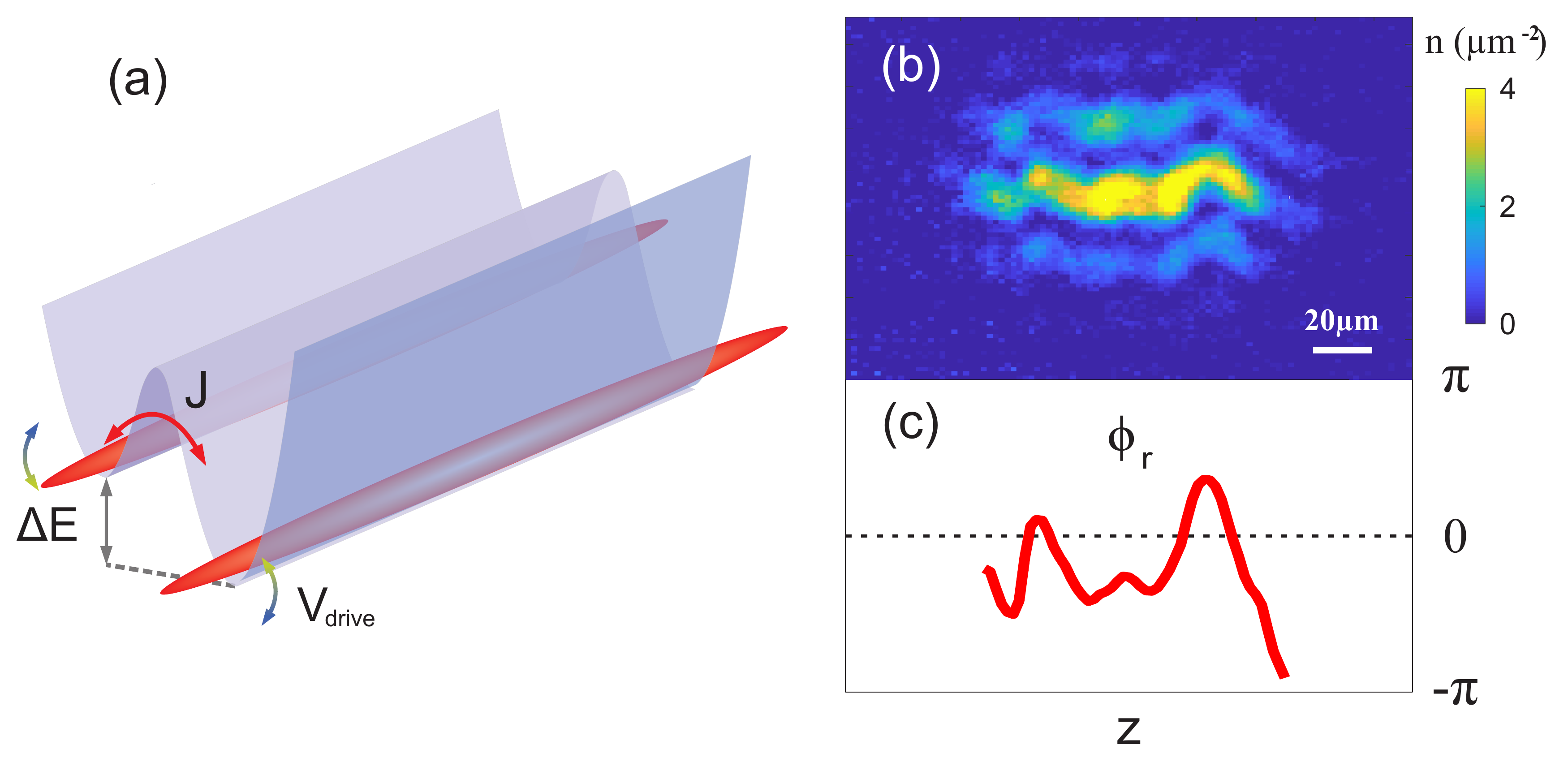}
\caption{\label{fig:Setup}
(a) Two atomic clouds in a tilted double-well potential, where $\Delta E$ is the energy difference between the two minima. For $\Delta E=0$ the barrier results in tunnel coupling $J$, completely suppressed for $\Delta E \gg \hbar \omega_J$ (see text). Floquet engineering is realized by periodically modulating the energy difference $\Delta E$ with an amplitude $V_\mathrm{drive}$. (b) Typical matter-wave interference pattern after $15.6\,\mathrm{ms}$ Time-of-Flight with atom density depicted in color and (c) the extracted relative phase $\phi_r (z)$ between the two superfluids.
}\end{figure}

\textit{Experimental setup and Floquet engineering}---
Our experiment (Fig.~\ref{fig:Setup}) starts with a pair of tunnel-coupled one-dimensional (1D) Bose gases of $^{87}$Rb atoms trapped in a DW potential created on an atom chip \cite{Fol00, Rei11} by radio-frequency (RF) dressing \cite{Hof06,Les06}. For both wells, the trapping frequencies are $\omega_\bot=2\pi \times 1.4$\,kHz transversely and $\omega_\mathrm{z}=2\pi \times 10$\,Hz longitudinally.
We prepare the system (total atom number $N \approx 10^4$, peak density $\rho_0 \approx 50$\,\textmu$\mathrm{m}^{-1}$) through evaporative cooling in a balanced DW potential. The initial temperature of the samples is $T_\text{i} \approx  37(5)\,\mathrm{nK}$ as measured by the two-point density correlation function after $11.2\,\mathrm{ms}$ Time-of-Flight (ToF) \cite{Man10}. 
In the longitudinal direction, the spatially resolved relative phase between the superfluids is extracted from matter-wave interference after $15.6\,\mathrm{ms}$ ToF. For the further analysis we use the central $50 \,$\textmu$\mathrm{ m}$ (variation of density less than 15\%). Expectation values are calculated by averaging over 30-60 experimental realizations.

The RF dressing allows precise control over the DW potential, determining the barrier height, i.e.~the tunnel coupling $J$, and the potential energy difference $\Delta E$ between the two wells (Fig.~\ref{fig:Setup}(a)), thereby realizing an extended Josephson junction (JJ). For $\Delta E=0$ such a system is a quantum simulator for the sine-Gordon (SG) quantum field theory \cite{Gri07,Sch17,SI}.
The Josephson frequency $\omega_J = \sqrt{4 J \mu / \hbar}$ is determined by the on-site interaction energy $\mu$ and the tunneling strength $J$ in the balanced (untilted) DW. Tunneling between the two superfluids can be completely suppressed by tilting the DW with an energy difference $\Delta E \gg \hbar \omega_J$. Floquet engineering enables us to revive the tunnel coupling within the tilted DW through a periodic modulation $\Delta E(t) = \Delta E_0 + 2 V_\mathrm{drive} \operatorname{sin}(\omega t + \varphi_\mathrm{drive})$. 

If the modulation frequency $\omega$ is near resonant with $\Delta E_0$, Floquet assisted tunneling will re-couple the two superfluids, even if the amplitude $V_\mathrm{drive} \ll \Delta E_0$. After time-averaging over this fast modulation, the system can effectively be treated as a static balanced DW. 
In the rotating Floquet frame, the time averaged Hamiltonian in the two-mode approximation is given by \cite{Gri98,SI}
\begin{equation} \label{eq:Ham}
\begin{aligned}
\langle\widetilde{H}(t)\rangle
=-\hbar \widetilde{J} \left( \begin{array}{cc}
0 &    \mathrm{e}^{-i(\varphi_\mathrm{drive}+\pi/2)}\\
 \mathrm{e}^{+i(\varphi_\mathrm{drive}+\pi/2)}    &  0 \end{array} \right ) ~,
\end{aligned}
\end{equation}
where $\widetilde{J}=J \times \mathcal{B}_1 (\frac{2V_\mathrm{drive}}{\hbar \omega})$ is the revived effective tunneling strength and $\mathcal{B}_1$ is the first order Bessel function. The ground state of the Hamiltonian \eqref{eq:Ham} takes the form
\begin{equation}\label{eq:groundstate}
\begin{aligned}
\lvert\widetilde{\Psi}_{\mathrm{ground}} \rangle=\frac{1}{\sqrt{2}}
\left(
\begin{array}{c}
\operatorname{exp}[-i (\frac{\varphi_\mathrm{drive}}{2}+\frac{\pi}{4})]\\
\operatorname{exp}[+i (\frac{\varphi_\mathrm{drive}}{2}+\frac{\pi}{4})]
\end{array}\right)\;  ~,
\end{aligned}
\end{equation}

The amplitude and phase of the tunneling term in the Floquet Hamiltonian Eq.~\eqref{eq:Ham} can be controlled through the amplitude $V_\mathrm{drive}$ and phase $\varphi_\mathrm{drive}$ of the modulation. For the ground state Eq.~\eqref{eq:groundstate}, the latter results in a non-zero relative phase $\widetilde{\phi}_\mathrm{r}=\varphi_\mathrm{drive}+\pi/2$ between the two superfluids when $\varphi_\mathrm{drive}\neq -\pi/2$. Transforming the relative phase $\widetilde{\phi}_\mathrm{r}$ in the rotating Floquet frame back to the laboratory frame, the measured relative phase between the two wells is given by
\begin{equation} \label{eq:phase_lab_frame}
\phi_\mathrm{r}=\widetilde{\phi}_\mathrm{r}+\Delta E_0 t/\hbar -\frac{2V_\mathrm{drive}}{\hbar \omega} \operatorname{cos}(\omega t+\varphi_\mathrm{drive}) ~.
\end{equation}
Hence, when $\varphi_\mathrm{drive}=-\pi/2$, the initial state with $\phi_{r}=0$ in lab frame is also the ground state with $\widetilde{\phi}_{r}=0$ in the Floquet frame. On the other hand, for an arbitrary modulation phase $\varphi_\mathrm{drive}$, 
the state $\phi_r=0$ is, in the Floquet frame, mapped to a JJ with a non-vanishing starting phase \mbox{$\Delta \widetilde{\phi}_{r,0}=\frac{2 V_\mathrm{drive}}{\hbar \omega} \operatorname{cos}(\varphi_\mathrm{drive})-\varphi_\mathrm{drive}-\pi/2$}.
For later convenience, we introduce $\Delta \widetilde{\phi}_r=\widetilde{\phi}_r-\varphi_\mathrm{drive}-\pi/2$, which shifts the ground state of the JJ back to zero phase.
The ability to tune the coupling strength and initial relative phase of such an extended bosonic JJ establishes the building block for more elaborate Floquet engineering.

\textit{Floquet assisted tunneling}.---
We first consider the simplest case, preparing the initial state as the ground state in the Floquet frame, i.e.~choosing the modulation phase $\varphi_\mathrm{drive}=-\pi/2$: We start in a strongly phase locked initial state with $\phi_\mathrm{r} \approx 0$, prepared by cooling into a strongly coupled, balanced DW. We then completely suppress the tunneling between the two wells rapidly by introducing an energy difference $\Delta E_0=h\times411\,\mathrm{Hz} \gg \hbar \omega_J$ and, at the same time, begin the periodic modulation $\Delta E(t)$. We choose the driving frequency $\omega=\Delta E_0 / \hbar$ resonant with the energy detuning and a driving amplitude $V_\mathrm{drive}=h \times 85\,\mathrm{Hz}$, leading to an expected revived tunneling strength of $\widetilde{J} \approx 0.2 J$. 

\begin{figure}
\includegraphics [scale=0.15] {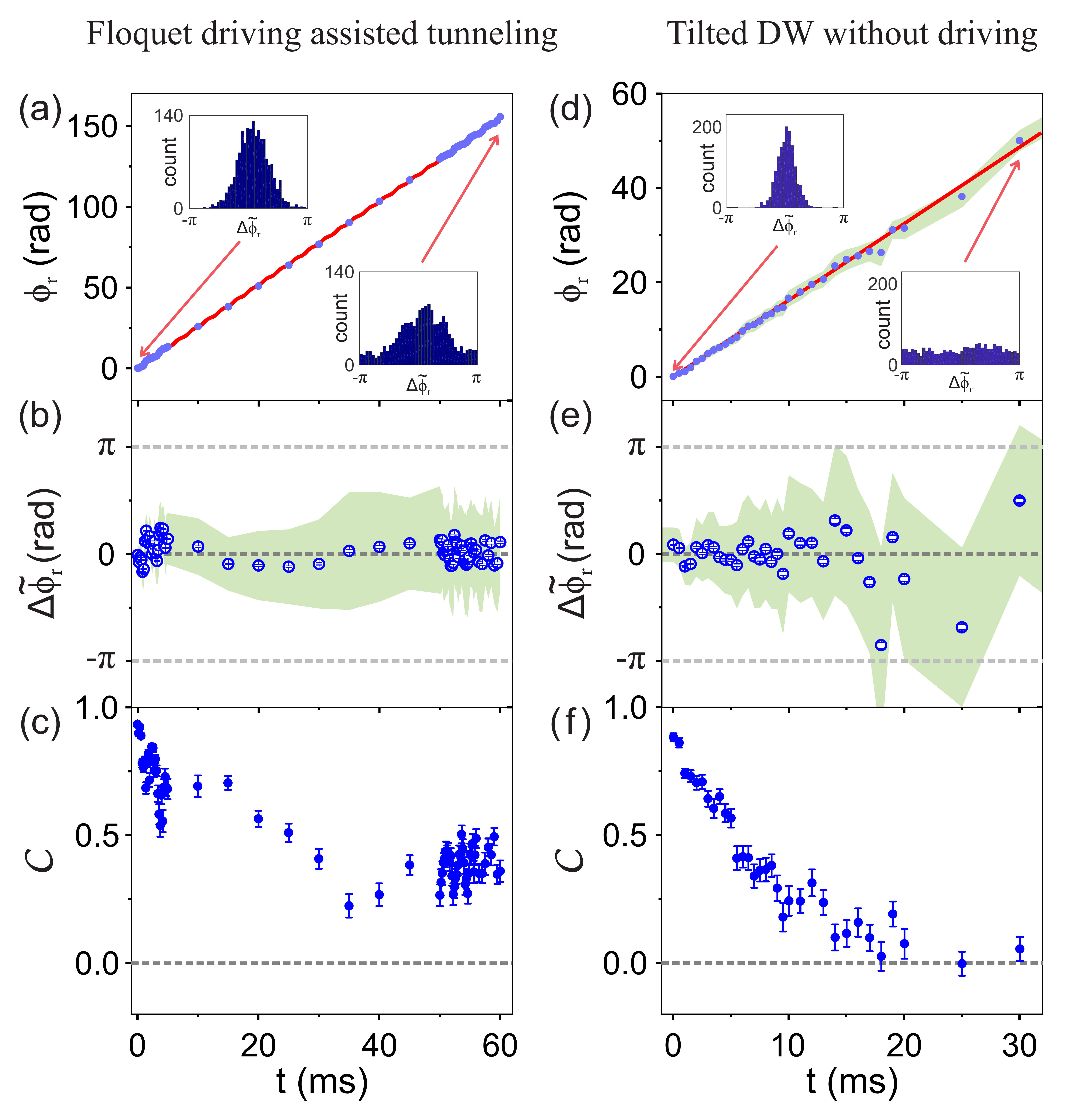}
\caption{\label{fig:Shaking_sample}
(left) Floquet driven DW: $\Delta E=h\times411\:$ Hz with Floquet drive $\omega=2\pi \times 411$\,Hz, starting phase $\varphi_\mathrm{drive}=-\pi/2$ and modulation amplitude $V_\mathrm{drive}=h\times 85\:$ Hz: (a) Evolution of the relative phase 
$\phi_\mathrm{r}$, calculated via the ensemble and spatially averaged circular mean of $\phi_\mathrm{r}(z)$ and shifted by $2 \pi \mathbb{N}$; (b) difference $\Delta\widetilde{\phi}_\mathrm{r}$ between the measured relative phase $\phi_\mathrm{r}$ and the estimate from Floquet theory; (c) Coherence factor. (right) (d-f) Comparison to a static tilted DW with $\Delta E=h\times258$\,Hz. Blue dots: experimental data. Red line: prediction from Floquet theory. The insets in (a) and (d) show the experimental distribution of $\Delta\widetilde{\phi}_\mathrm{r}$(z) and the light green region in (b), (d) and (e) display its standard deviation. The error bars give the standard error of the mean.}
\end{figure}

In Figure~\ref{fig:Shaking_sample} we present such a Floquet assisted tunneling experiment (left column) and compare it to the static tilted DW (right column). Figure~\ref{fig:Shaking_sample}(a) shows the time evolution for the spatially averaged relative phases (blue dots), in good agreement with the theoretical predictions (red line) given by Eq.~\eqref{eq:phase_lab_frame} with $\widetilde{\phi}_\mathrm{r}=0$. The insets show the full distribution functions of $\widetilde{\phi}_\mathrm{r}$ for the initial and final states. The observed broadening of the distribution reflects the relaxation in the SG model following a quench to a lower coupling $\widetilde{J}$. The small, random deviations $\Delta\widetilde{\phi}_r$ (Fig.~\ref{fig:Shaking_sample}(b)) represent the fluctuations of the field within the Floquet engineered SG Hamiltonian.

The revived tunneling strength $\widetilde{J}$ can be quantified via the coherence factor $\mathcal{C}=\langle \operatorname{cos}(\widetilde{\phi}_\mathrm{r}) \rangle$, depicted in Fig.~\ref{fig:Shaking_sample}(c). The initial state with $\phi_\mathrm{r} \approx 0$ shows almost perfect phase coherence $\mathcal{C} >0.9$, whereas for $\mathcal{C} \approx 0$ the relative phase is completely random. 
At early times, $t \lesssim 5\,\mathrm{ms}$, we observe a fast decrease of coherence due to quasi-particle dephasing caused by the quench to a smaller effective tunneling coupling $\widetilde{J}$.
For $t \gtrsim 30\,\mathrm{ms}$ the system reaches a quasi-steady state with an average coherence factor $\mathcal{C} \approx 0.38$ (Fig.~\ref{fig:Shaking_sample}(c)). This demonstrates that the system retains finite coherence due to a non-vanishing Floquet assisted tunneling coupling, even after relaxation.

For comparison we show in Fig.~\ref{fig:Shaking_sample}(d)-(f) the evolution for a static tilted DW with $\Delta E_0 = h \times 258\,\mathrm{Hz}$ and no periodic modulation (i.e.~$V_\mathrm{drive}=0$). The evolution of the relative phase within the first $10\,\mathrm{ms}$ is in good agreement with Eq.~\eqref{eq:phase_lab_frame}. The constant energy difference $\Delta E_0$ only leads to a monotonic accumulation of a relative phase. For $t \gtrsim 20\,\mathrm{ms}$ the total suppression of the initial tunneling leads to a fully random phase distribution (Fig.~\ref{fig:Shaking_sample}(d), inset) and consequently a vanishing coherence factor $\mathcal{C} \approx 0$. 
The detuning $\Delta E_0$ was chosen to be the minimal energy difference between the two wells in the periodically modulated case. Consequently $\mathcal{C} \to 0$ in Fig.~\ref{fig:Shaking_sample}(f) verifies the complete suppression of static coupling in the Floquet system at any time and $\mathcal{C} \gg 0$ in Fig.~\ref{fig:Shaking_sample}(c) can only come from tunneling revived through Floquet engineering.
 
In the next experiment, we demonstrate that Floquet engineering can be used to recouple two independent superfluids, leading to the build up of coherence between the two wells. As before, we prepare the system in a balanced DW, but now with a high barrier and an initial coupling $J \approx 0$. Hence, the initial state after cooling is completely uncorrelated, i.e.~the relative phase distribution is fully random and $\mathcal{C} \approx 0$. We subsequently lower the barrier height between the two wells while at the same time suppressing the coupling $J$ by introducing an energy difference $\Delta E_0 = h \times 380\,\mathrm{Hz}$. The reshaping of the DW potential is done within $5\,\mathrm{ms}$, short  compared to the tunneling time ($\sim 2 \pi/\omega_J$) but sufficiently long to avoid radial excitations ($\sim 2 \pi/\omega_\bot$). Thereafter, we proceed with our usual Floquet modulation with frequency $\omega = \Delta E_0 / \hbar$ and amplitude $V_\mathrm{drive}=h \times 190\,\mathrm{Hz}$ which recouples the DW.

\begin{figure}
\includegraphics[scale=0.155]{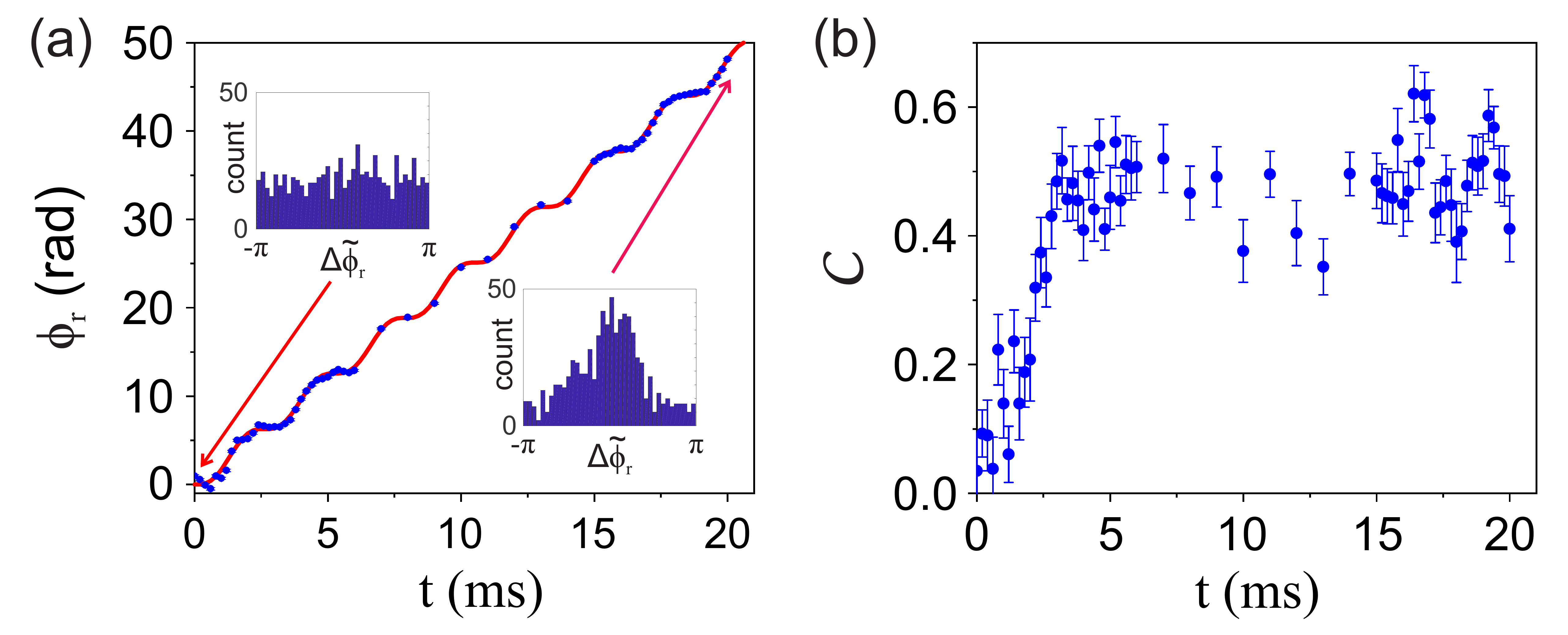}
\caption{\label{fig:recouple} 
The evolution of (a) relative phase $\phi_r$ and (b) coherence factor $\mathcal{C}$ of two initially uncorrelated Bose gases after switching on Floquet assisted tunneling. The driving is resonant with the energy difference of the DW: $\omega=2\pi \times 380\,\mathrm{Hz}$ and its amplitude is: $V_\mathrm{drive}=h\times 190\,\mathrm{Hz}$.
}
\end{figure}

Figure~\ref{fig:recouple} shows the time evolution of the relative phase $\phi_r$ and the coherence factor $\mathcal{C}$ for such a Floquet recoupling experiment. The evolution of the relative phase $\phi_\mathrm{r}$ shows good agreement with the theoretical predictions. The initial distribution of $\widetilde{\phi}_\mathrm{r}$ is uniform but rapidly narrows around zero (see insets), illustrating the buildup of coherence between the two wells.
This is also clearly visible in the increase of the coherence factor $\mathcal{C}$ depicted in Fig.~\ref{fig:recouple}(b). Within the first $5\,\mathrm{ms}$, corresponding to only 2 periods of the modulation, $\mathcal{C}$ increases to its plateau value $\mathcal{C} \approx 0.5$. This demonstrates fast phase locking by tunneling restored through Floquet engineering.

\begin{figure}
\includegraphics[scale=0.11]{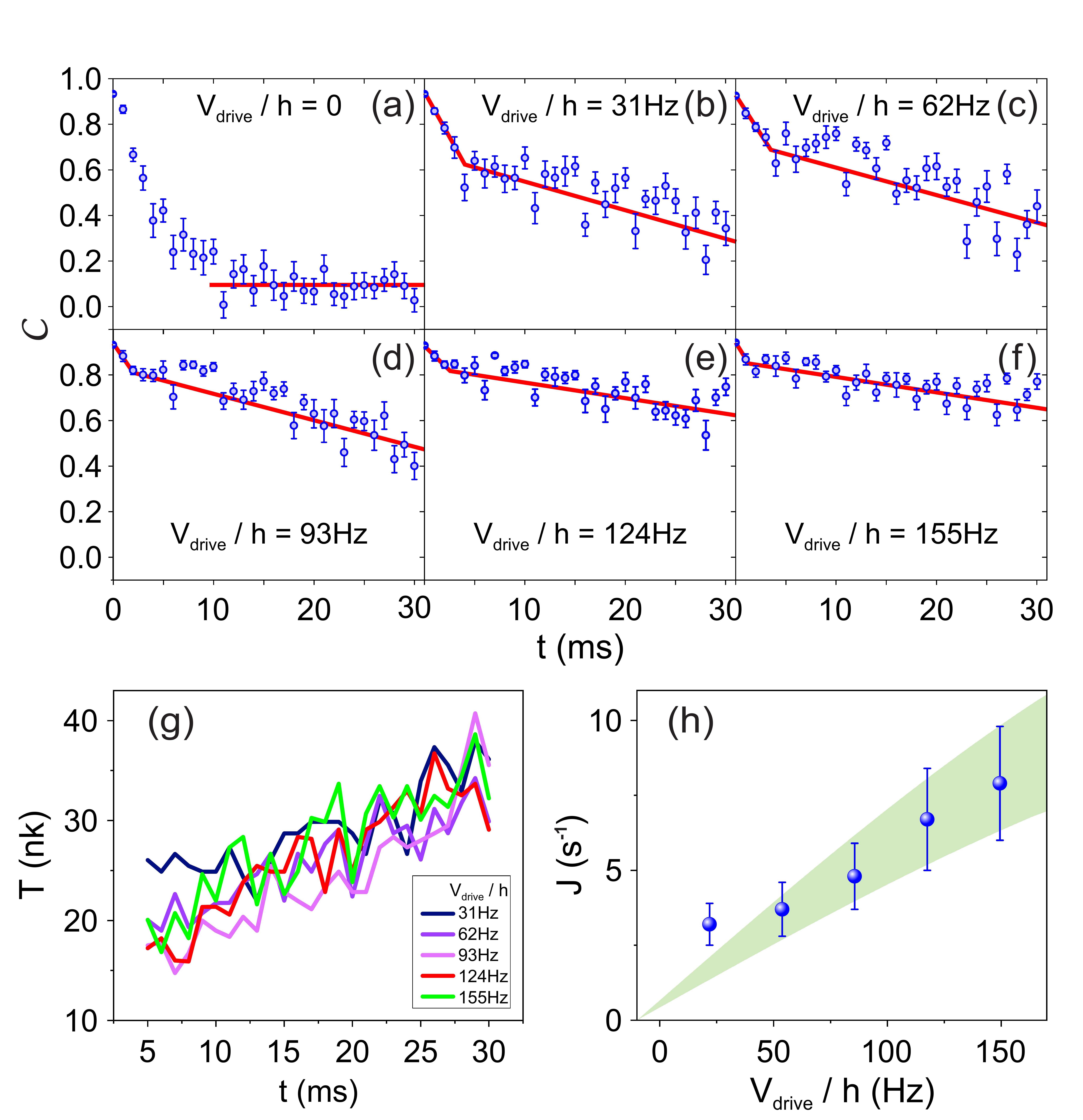}
\caption{\label{fig:DifferentAmp} 
(a-f). The evolution of the coherence factor $\mathcal{C}$ under Floquet driving with different amplitudes. The red lines are linear fits highlighting the two different time regimes discussed in the text. (g) The heating rates for the driven systems in (b-f). (h) Dependence of the Floquet induced Josephson tunneling on the driving amplitude. Blue dots: experiment; light green region: estimation from Floquet theory.} 
\end{figure}

\textit{Tunneling strength and heating}.---
Having established Floquet engineered tunneling, we proceed to a detailed experimental analysis of the tunneling strength $\widetilde{J}$ and the long time evolution of the system. 
Figure~\ref{fig:DifferentAmp}(a-f) shows the time evolution of the coherence factor $\mathcal{C}$ for six different driving amplitudes $V_\mathrm{drive}/h$ ranging from 0 to $155\,\mathrm{Hz}$. The experimental sequence and all other parameters are the same as for Fig.~\ref{fig:Shaking_sample}(a-c). Qualitatively, the increase of the Floquet engineered coupling $\widetilde{J}$ is readily apparent from the increase of $\mathcal{C}$ with the amplitude $V_\mathrm{drive}$. 

The time evolution shown in Fig.~\ref{fig:DifferentAmp}(b-f) can clearly be divided into two stages: 
Within the first $5\, \mathrm{ms}$, the quench of the tunneling coupling from ${J}$ to  $\widetilde{J} < J$ leads to quasi-particle dephasing and results in a rapid decline of $\mathcal{C}$. This is consistent with previous observations for the relaxation following a quench in static DW potentials \cite{Hof07,Gring2012,Sch21}. In the second stage, $t \gtrsim 5 \,\mathrm{ms}$, we observe a further slow decline of coherence due to heating. 

We self-consistently extract the Floquet tunneling coupling $\widetilde{J}$ and the time dependent temperature (Fig.~\ref{fig:DifferentAmp}(g)) by comparing the measured correlation functions after dephasing to SG model predictions in thermal equilibrium \cite{Gri13,Sch17,Beck18} under the assumption that the Floquet tunneling strength $\widetilde{J}$ stays constant during the driving. Remarkably the derived heating rate $\Gamma \sim 0.6 \, \mathrm{nK/}$\textmu$\mathrm{s}$ is rather independent of the driving amplitude $V_\mathrm{drive}$ 
(see \ref{Tab:Data}).
%(table 1 in \cite{SI}).

Equivalently, using the initial temperature and correlation functions we determine the initial tunneling coupling $J = 23\pm5 \,s^{-1}$ in the initial balanced DW. In Fig.~\ref{fig:DifferentAmp}(h) we compare the measured $\widetilde{J}$ to the calculated Floquet tunneling strengths Eq.~\eqref{eq:Ham}. We find good agreement between experimental results and Floquet theory.

The common mode temperature of the system before and after $30\,\mathrm{ms}$ of Floquet driving was measured by density correlation functions after $11.2\,\mathrm{ms}$ ToF \cite{Man10}. We found no significant heating in the common mode (see \ref{Tab:Data} and \cite{SI} for details).
%(table 1 in \cite{SI}). 

\textit{Josephson oscillations and relaxation}.---
Finally we consider different driving  phases $\varphi_\mathrm{drive}$ of the modulation, which imprints an initial relative phase difference $\Delta \widetilde{\phi}_{\mathrm{r},0}$ directly in the Floquet frame. This realizes an extended JJ and permits precise control over the initial conditions. 
In the experiments we implement the same experimental Floquet sequence as before, starting from a strongly tunneling coupled state with $\phi_\mathrm{r} \approx 0$ and $\mathcal{C} \approx 0.9$,  detuning $\Delta E_0 = h \times 436\, \mathrm{Hz}$, and driving amplitude $V_\mathrm{drive} = h \times 93\, \mathrm{Hz}$, but now considering different driving phases $\varphi_\mathrm{drive} = 0, \, \pm \pi/6, \, \pm \pi/3, \, \pm \pi/2$. Using Eq.~\eqref{eq:phase_lab_frame}, we get the initial phase differences $\Delta \widetilde{\phi}_\mathrm{r,0} \in [-\pi,0]$. 

\begin{figure}
\includegraphics[scale=0.21]{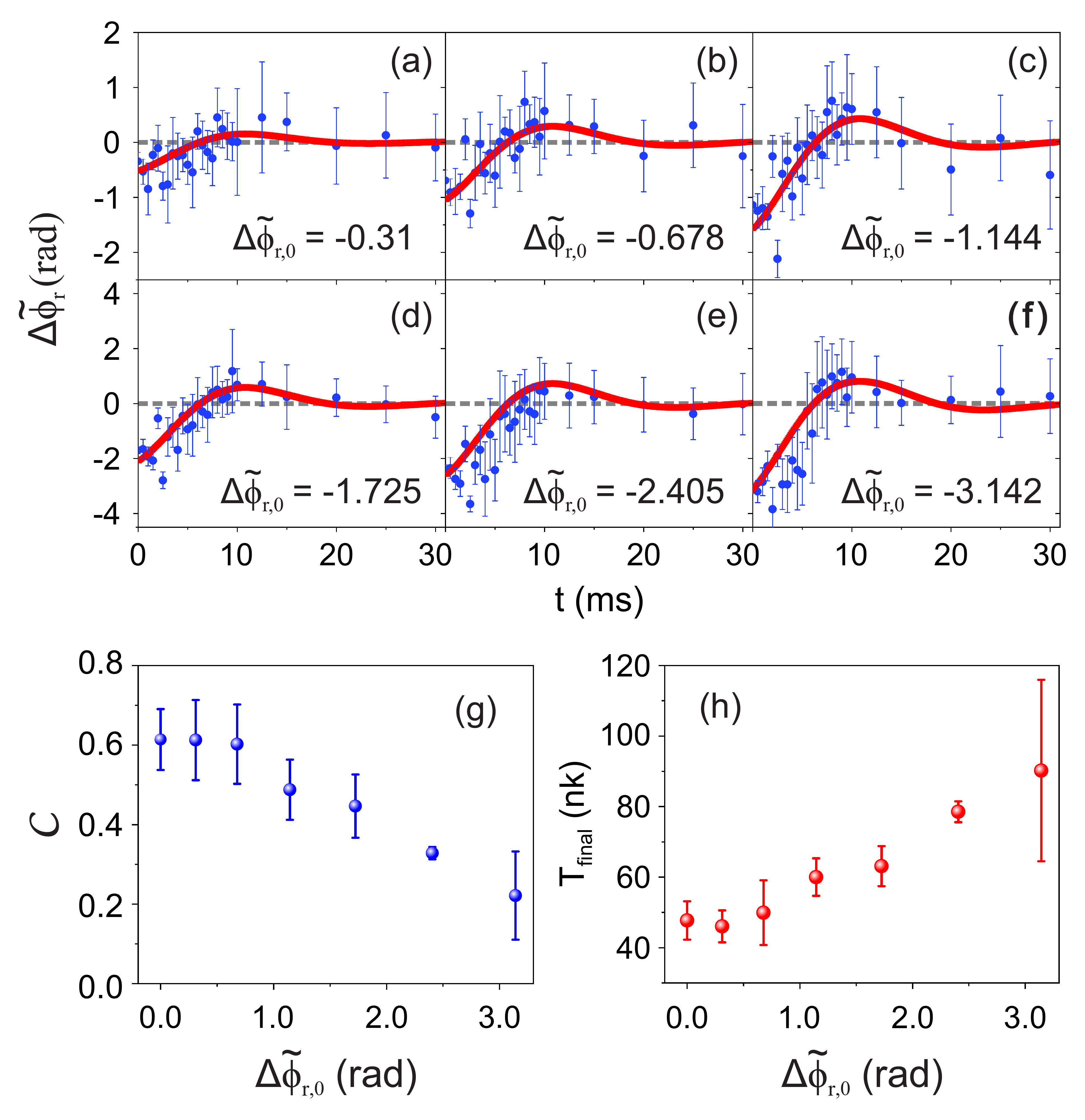}
\caption{\label{fig:DifferentPhase} 
(a-f) Josephson oscillations starting from different initial relative phases $\Delta\widetilde{\phi}_{r,0}$ in the rotating Floquet frame. The red line is fitted with the assumption that the Josephson frequencies $\omega_J$ and the decay rates $\tau$ are independent of the initial phase $\Delta\widetilde{\phi}_{r,0}$. (g) Coherence factor and (h) temperature of the system after $20\,\mathrm{ms}$ of driving for different initial relative phase $\Delta\widetilde{\phi}_{r,0}$. }
\end{figure}

Figure~\ref{fig:DifferentPhase}(a-f) shows the time evolution of the relative phase for different $\varphi_\mathrm{drive}$. 
In all cases we find strongly damped Josephson oscillations, in accordance with previous experiments in static DW potentials \cite{Pig18}.

We quantify the evolution by fitting a damped oscillation in order to extract the Josephson frequency $\omega_J$ and the characteristic damping time $\tau$. From the previous section, we expect a Floquet tunneling strength $\widetilde{J} \approx 4.8\, s^{-1}$ and $\omega_J \approx 2 \pi \times 46\,\mathrm{Hz}$, independent of $\varphi_\mathrm{drive}$. We therefore fit the experimental data with a single Josephson frequency and damping time, leading to $\omega_J^\mathrm{fit} = 2\pi\times40(3)\,\mathrm{Hz}$ and $\tau^\mathrm{fit} = 8.8(1.2)\,\mathrm{ms}$ respectively. Note that $\omega_J^\mathrm{fit}$ agrees with the theoretical expectations within the statistical error, thus being in good agreement with the exact Floquet predictions for $\widetilde{J}$. The damping time $\tau^\mathrm{fit}$ is compatible with damping times observed in static DW potentials \cite{Pig18,Men21}.

We consistently observe relaxation to a stationary state after $t \gtrsim 20\,\mathrm{ms}$. This indicates transfer of the initial potential energy, driving the coherent Josephson oscillation, into fluctuations of the relative phase \cite{Sch19}. In Fig.~\ref{fig:DifferentPhase}(g,h) we show the time-averaged coherence factor $\mathcal{C}$ and final temperature of the system for $t > 20\,\mathrm{ms}$, respectively. The decrease (increase) of coherence (temperature) with $\Delta \widetilde{\phi}_{\mathrm{r},0}$ reflects the increased potential energy $\sim \widetilde{J} \operatorname{cos}(\Delta \widetilde{\phi}_\mathrm{r,0})$ of the initial state. Notably, we again find the common mode temperatures to remain approximately constant throughout the evolution. 

\textit{Conclusion}.---
Floquet engineering allows to revive the tunnel coupling in a tilted double well, creating an extended bosonic Josephson junction. We find excellent quantitative agreement between the experimental results and theoretical predictions. The periodic modulation technique developed in this work will greatly expand the freedom of manipulating such double well systems and open up new ways to study many intriguing fundamental phenomena with broad relevance ranging from condensed matter physics to cosmology.

\begin{acknowledgments}
We thank I. Mazets, B. Rauer and M. Serbyn for helpful discussions. This work is supported by the DFG/FWF Collaborative Research Centre `SFB 1225 (ISOQUANT)', and by the Wiener Wissenschafts- und TechnologieFonds (WWTF), Project No.MA16-066 (SEQUEX). F.C., F.S.M., and J.~Sabino acknowledge support by the Austrian Science Fund (FWF) in the framework of the Doctoral School on Complex Quantum Systems (CoQuS). T.S. acknowledges support from the Max Kade Foundation through a postdoctoral fellowship. S.-C.J. and S.E. acknowledges support through an ESQ (Erwin Schr\"odinger Center for Quantum Science and Technology) fellowship funded through the European Union’s Horizon 2020 research and innovation program under Marie Skłodowska-Curie Grant Agreement No 801110. This project reflects only the author’s view, the EU Agency is not responsible for any use that may be made of the information it contains. ESQ has received funding from the Austrian Federal Ministry of Education, Science and Research (BMBWF). J. Sabino acknowledges support by the Funda\c{c}\~{a}o para a Ci\^{e}ncia e a Technologia (PD/BD/128641/2017).
\end{acknowledgments}

\bibliography{Shaking_tunneling_citation}

\vspace{8mm}
\appendix

\setcounter{figure}{0}
\setcounter{equation}{0}
\renewcommand{\theequation}{S\arabic{equation}}
\renewcommand{\thefigure}{S\arabic{figure}}
\renewcommand{\thetable}{S\arabic{table}}

{
\centering
\textbf{SUPPLEMENTAL MATERIAL} \\
}
\vspace{4mm}

\begin{table}[h]
\begin{tabular}{ |p{1.73cm}||p{1.35cm}|p{1.29cm}|p{1cm}|p{1.42cm}| p{1cm}|}
 \hline
 $V_\mathrm{drive}/h$ \!\!(Hz) & $J_\mathrm{exp}$ \!\!($s^{-1}$) & $J_\mathrm{Flo}$ \!\!($s^{-1}$) & $T_\mathrm{r}^f$ \!\!\!(nK) &$\Gamma$ \!\!(nK/ms) &$T_\mathrm{c}^f$ \!\!\!(nK)\\
 \hline
31     &  3.2(0.7)  & 1.7(0.4)    &  42(3)    & 0.50 & 35(3) \\
62     & 3.7(0.9)   & 3.4(0.7)    &  32(2)   & 0.50 & 34(5) \\
93     & 4.8(1.1)   & 5.0(1.1)   &  37(3)    & 0.60 & 26(4) \\
124    & 6.7(1.7)   & 6.6(1.4)    &  32(2)    & 0.69 & 28(5)  \\
155    & 7.9(1.9)   & 8.1(1.7)     &  35(3)    & 0.64 & 38(5)  \\
 \hline
\end{tabular}
\caption{\label{Tab:Data} Results for different modulation amplitudes $V_\mathrm{drive}$ corresponding to Fig.~\ref{fig:DifferentAmp}. Depicted values are the experimentally measured ($J_\mathrm{exp}$) and theoretically expected ($J_\mathrm{Flo}$, see Eq.~\eqref{eq:coupling_Floquet}) tunneling coupling, the heating rate ($\Gamma$), and the final temperature for both the relative ($T_\mathrm{r}^f$) and common ($T_\mathrm{c}^f$) degrees of freedom.
}
\end{table}
\vspace{-8mm}

\section{Floquet Hamiltonian engineering of DW system}

The derivation of the Floquet Hamiltonian for our DW system is based on the single-particle picture
\begin{align} \label{eq:Ham1}
\widehat{H}(t)&=\widehat{H}_0+\widehat{H}_t \\
&=\left( \begin{array}{cc}
\Delta E_0 & -\hbar J\\
-\hbar J       &  0 \end{array} \right )+
V_\mathrm{drive} \sin (\omega t+\varphi_\mathrm{drive}) \cdot  \sigma_\mathrm{z} ~, \nonumber
\end{align}
where $J$ is the tunneling strength in the balanced double well, $\Delta E_0$ is the energy difference between the two trap bottoms, $\omega$ is the driving frequency, $\varphi_\mathrm{drive}$ is the starting phase of the Floquet modulation, and $\sigma_\mathrm{z}$ is the Pauli matrix.

To simplify this Hamiltonian, we perform a unitary transformation to the Floquet frame
\begin{equation}
\widetilde{H}(t)=\widehat{R}(t)\widehat{H}(t)\widehat{R}^\dagger(t)-i\hbar\widehat{R}(t) \frac{d\widehat{R}^\dagger(t)}{dt} ~,
\end{equation}
with the unitary operator
\begin{equation} \label{eq:Rotation}
\widehat{R}(t)=\left( \begin{array}{cc}
e^{i \frac{\Delta E_0}{\hbar} t- i \mathcal{V}(t)} & 0 \\
 0     &  e^{+i \mathcal{V}(t)} \end{array} \right ) ~.
\end{equation}
Here we defined 
\begin{equation}
\mathcal{V}(t) = \frac{V_\mathrm{drive}}{\hbar \omega}\cos(\omega t+ \varphi_\mathrm{drive})
\end{equation}
to shorten the notation. This eliminates the diagonal term in Eq. (\ref{eq:Ham1}) leading to:
\begin{equation} \label{eq:Ham2}
\widetilde{H}(t)= -\hbar J \left( \begin{array}{cc}
0 &  \mathrm{e}^{+i \frac{\Delta E_0}{\hbar} t- i 2 \mathcal{V}(t)}\\
  \mathrm{e}^{-i \frac{\Delta E_0}{\hbar} t+ i 2 \mathcal{V}(t)}      &  0 \end{array} \right ) ~.
\end{equation}

Based on the Floquet theory, we can achieve the effective Hamiltonian by calculating the time-average of Eq.~\eqref{eq:Ham2} in one modulation period. When the driving frequency is resonant with the energy difference, i.e. $\hbar \omega=\Delta E_0$, the time-averaged Floquet Hamiltonian is given by
\begin{equation} \label{eq:Ham3}
\langle\widetilde{H}(t)\rangle=-\hbar \widetilde{J} \left( \begin{array}{cc}
0 &   e^{-i(\varphi_\mathrm{drive}+\pi/2)}\\
e^{+i(\varphi_\mathrm{drive}+\pi/2)}    &  0 \end{array} \right ) ~,
\end{equation}
where 
\begin{equation} \label{eq:coupling_Floquet}
\widetilde{J}=J\cdot \mathcal{B}_1(\frac{2V_\mathrm{drive}}{\hbar \omega})
\end{equation}
is the effective tunneling strength determined by the first order Bessel function \mbox{$\mathcal{B}_1(x)=\frac{1}{2 \pi} \int^{\pi}_{-\pi} \mathrm{d}\tau \, e^{i(x\sin\tau-\tau)}$}. From Eq.~\eqref{eq:Ham3}, the eigenstates and corresponding eigenvalue can be easily solved:
\begin{eqnarray}
\lvert \widetilde{\Psi}_\pm \rangle&=&\frac{1}{\sqrt{2}}
\left(
\begin{array}{c}
\operatorname{exp}[-i (\frac{\varphi_\mathrm{drive}}{2}+\frac{\pi}{4})]\\
\pm \operatorname{exp}[+i (\frac{\varphi_\mathrm{drive}}{2}+\frac{\pi}{4})]
\end{array}\right)\; \label{eq:eigenstate}
\\
\varepsilon_\pm&=&\mp \hbar \widetilde{J}
\end{eqnarray}

As discussed in the main text, the relative phase of the eigenstate in the Floquet frame depends on the driving phase $\varphi_\mathrm{drive}$, which can be controlled experimentally through the RF dressing of the DW potential. The time evolution for an arbitrary state in the Floquet frame can readily be calculated from Eq. (\ref{eq:Ham3}). In order to compare with the results measured in the experiment, we calculate the relative phase evolution of $\lvert \widetilde{\Psi}_+\rangle$ (red line in Fig.~\ref{fig:Shaking_sample}(a) and in Fig.~\ref{fig:recouple}(a)) and afterwards transform $\widetilde{\phi}_r$ from the Floquet frame back to the laboratory frame via Eq.~\eqref{eq:phase_lab_frame}.

\section{Sine-Gordon model for tunnel-coupled superfluids}

For a static, balanced DW potential the sine-Gordon model was proposed \cite{Gri07} and in thermal equilibrium experimentally verified \cite{Sch17} to be the low energy effective description of a pair of tunneling coupled quasicondensates. 
A complete derivation of the Floquet engineered interacting many-body system would go far beyond the current paper. For the sake of completeness we give here a brief derivation for our static DW system, assumed to describe the time-independent Floquet Hamiltonian.

For the temperatures and atom numbers considered, both condensates, located in the left and right minimum of the DW potential, fulfill the 1D condition $\mu, k_\mathrm{B} T \ll \hbar \omega_\perp$. Since dynamics along the radial directions are frozen out, we can proceed with the common dimensional reduction by integrating over the tightly confined radial directions. Tunneling through the DW barrier couples the two quantum wires, leading to the effective one-dimensional Hamiltonian
\begin{align} \label{eq:1D-Ham}
H=&\sum_{j=1}^2 \int dz \Bigg[\frac{\hbar^2}{2m}\frac{\partial \psi_j^{\dagger}}{\partial z} \frac{\partial \psi_j}{\partial z}+\frac{g_\mathrm{1D}}{2}\partial \psi_j^{\dagger}
\partial\psi_j^{\dagger} \partial\psi_j \partial \psi_j \nonumber \\
+&U(z)\psi_j^{\dagger} \psi_j \Bigg]
-\hbar J \int dz\Bigg[\psi_1^{\dagger}\psi_2+\psi_2^{\dagger} \psi_1 \Bigg] ~.
\end{align}
Here $m$ is the atomic mass, $g_\mathrm{1D}=2\hbar a_\mathrm{s} \omega_{\perp}$ is the 1D effective interaction strength, $a_\mathrm{s}$ is the s-wave scattering length, $U$ is the longitudinal potential, and $2 \hbar J$ is the single particle tunneling-coupling energy. The field operators fulfill the bosonic commutation relative $[\psi_j(z), \psi_{j'}^{\dagger}(z')]=\delta_{jj'}\delta(z-z')$.

\begin{figure}
\includegraphics[scale=0.21]{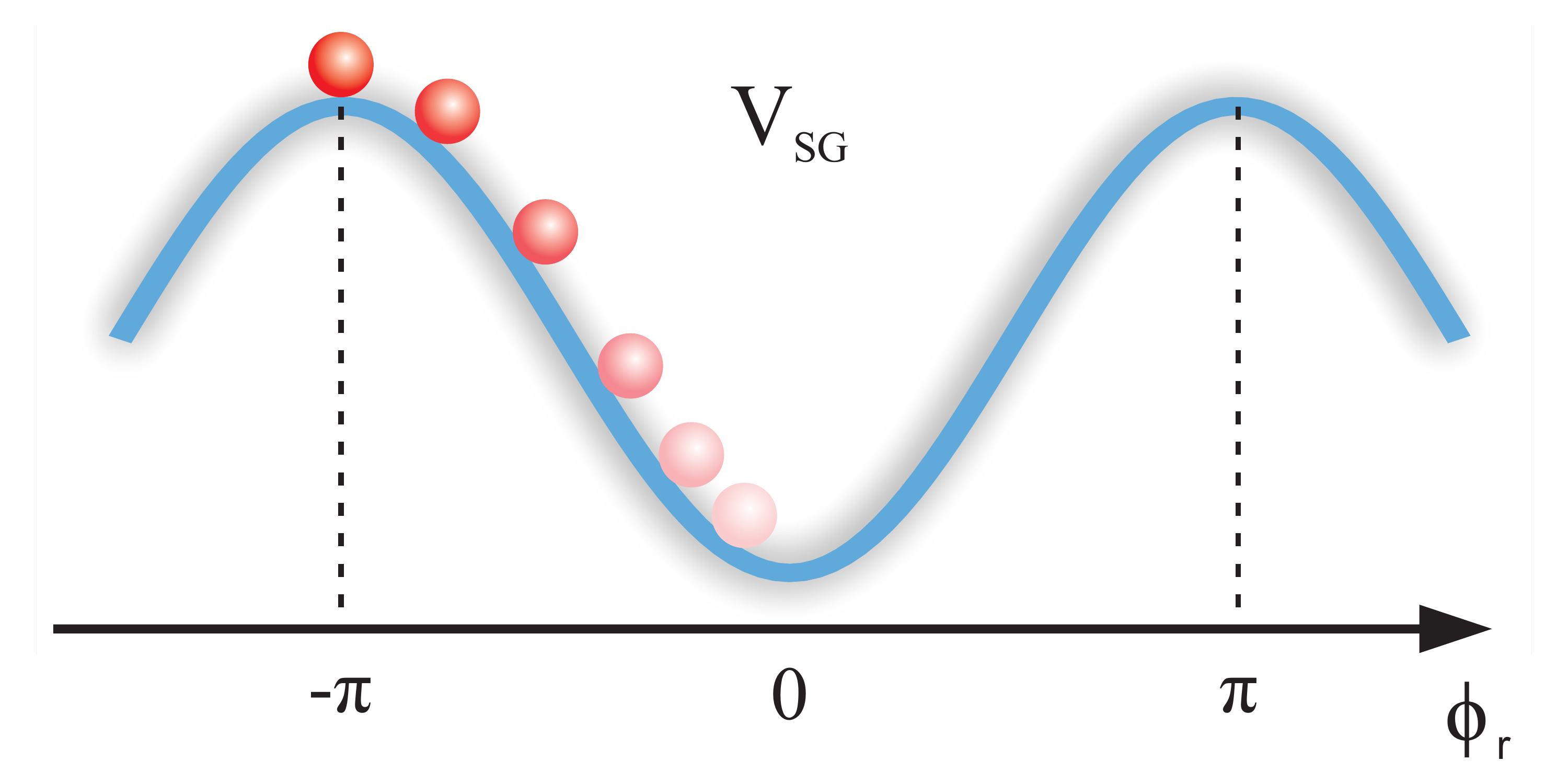}
\caption{\label{fig:S2} Schematic of the potential energy for the sine-Gordon Hamiltonian \eqref{eq:H_rel_SG}. The red spheres show the initial states for the different Floquet driving phases in Fig.~\ref{fig:DifferentPhase}(a)-(f). Note, that $\phi_\mathrm{r}$ here corresponds to the shifted relative phase $\Delta \widetilde{\phi}_\mathrm{r}$, defined in the main text.}
\end{figure}

Expressing the wave function in terms of density and phase fluctuations 
\begin{equation} \label{eq:density-phase}
\psi_j(z)=\mathrm{exp}[i\theta_j(z)]\sqrt{n_\mathrm{1D}+\delta \rho_j(z)} ~,
\end{equation}
with canonical commutators $[\delta \rho_j(z), \theta_{j'}(z')]=i\delta_{jj'}\delta(z-z')$, the low-energy effective theory can be derived by expanding the Hamiltonian \eqref{eq:1D-Ham} to second order in the small density perturbations $\delta \rho_j$ and phase gradients $\partial_z \theta_j$. This separates the Hamiltonian \eqref{eq:1D-Ham} in a weakly coupled sum $H = H_s + H_r + V_{c,r}$ for the common (symmetric, 's') and relative (anti-symmetric, 'r') degrees of freedom (DoF), defined as
\begin{align} %\label{eq:symm-antisymm}
\delta \rho_s(z)=\delta \rho_1(z)+\delta \rho_2(z)~,~\phi_s(z)=\frac{1}{2}[\theta_1(z)+\theta_2(z)]~, \nonumber\\
\delta \rho_r(z)=\frac{1}{2}[\delta \rho_1(z)-\delta \rho_2(z)]~,~\phi_r(z)=\theta_1(z)-\theta_2(z) ~. \nonumber
\end{align}

Experiments in static DW potentials showed that in thermal equilibrium the coupling $V_{c,r}$ is negligible for a wide range of parameters \cite{Sch17}, such that the common and relative DoF are described by the Luttinger-Liquid and sine-Gordon Hamiltonian, respectively. The latter is given by
\begin{align} \label{eq:H_rel_SG}
H_\mathrm{r} = \! \int \! dz \Big[ g\delta \rho_\mathrm{r}^2 +\frac{\hbar^2 n_\mathrm{1D}}{4m} \left(\partial_z \phi_\mathrm{r}\right)^2 -2 \hbar J n_\mathrm{1D} \operatorname{cos}(\phi_\mathrm{r}) \Big] \,,
\end{align}
where, for simplicity, we consider the long wavelength limit (i.e.~neglecting derivatives of the density fluctuations). The first two terms represent the Luttinger-Liquid Hamiltonian describing massless phononic excitations. A schematic of the SG potential is depicted in Fig.~\ref{fig:S2}, including the initial states considered in Fig.~\ref{fig:DifferentPhase}(a)-(f). Their initial potential energy is transferred to fluctuations of the SG field, leading to the relaxation of the coherent Josephson oscillation in this extended bosonic JJ.

\end{document}